\documentclass[sigconf]{acmart}
\usepackage{booktabs} % For formal tables

\usepackage{xcolor}
\usepackage{graphicx}
\usepackage{subcaption}
\usepackage{soul}
\sethlcolor{lightgray}
\copyrightyear{2020}
\acmYear{2020}
\setcopyright{acmlicensed}\acmConference[SIGIR '20]{Proceedings of the 43rd International ACM SIGIR Conference on Research and Development in Information Retrieval}{July 25--30, 2020}{Virtual Event, China}
\acmBooktitle{Proceedings of the 43rd International ACM SIGIR Conference on Research and Development in Information Retrieval (SIGIR '20), July 25--30, 2020, Virtual Event, China}
\acmPrice{15.00}
\acmDOI{10.1145/3397271.3401221}
\acmISBN{978-1-4503-8016-4/20/07}

\begin{document}
\fancyhead{}
\title{Factuality Checking in News Headlines with Eye Tracking}
%\subtitle{SIGIR Long Paper (9 pages + 1 for refs) Deadline: 28 January (Abstract due 21 January)}
%\titlenote{Produces the permission block, and copyright information}
%\subtitle{Extended Abstract}
%\subtitlenote{The full version of the author's guide is available as \texttt{acmart.pdf} document}

\author{Christian Hansen}
\affiliation{
  \city{University of Copenhagen}
}
\email{chrh@di.ku.dk}

\author{Casper Hansen}
\affiliation{
  \city{University of Copenhagen}
}
\email{c.hansen@di.ku.dk}

\author{Jakob Grue Simonsen}
\affiliation{
  \city{University of Copenhagen}
}
\email{simonsen@di.ku.dk}

\author{Birger Larsen}
\affiliation{
  \city{Aalborg University}
}
\email{birger@hum.aau.dk}

\author{Stephen Alstrup}
\affiliation{
  \city{University of Copenhagen}
}
\email{s.alstrup@di.ku.dk}

\author{Christina Lioma}
\affiliation{
  \city{University of Copenhagen}
}
\email{c.lioma@di.ku.dk}

\begin{abstract}
We study whether it is possible to infer if a news headline is true or false using only the movement of the human eyes when reading news headlines. Our study with 55 participants who are eye-tracked when reading 108 news headlines (72 true, 36 false) shows that false headlines receive statistically significantly less visual attention than true headlines. 
We further build an ensemble learner that predicts news headline factuality using only eye-tracking measurements. Our model yields a mean AUC of 0.688 and is better at detecting false than true headlines. Through a model analysis, we find that eye-tracking 25 users when reading 3-6 headlines is sufficient for our ensemble learner.

%This is the first study of eye tracking for fact checking news headlines. 

%assigned  (iii) the manual labelling of unseen headlines as true or false by our human participants yields accuracy = ly inferring eye-tracked signals alone can be used to infer the label (true or fake) of a news item with a mean AUC of 0.678, which is higher than random, but lower than the mean AUC for labels created (consciously) by an emsemble of humans (AUC 0.937). These two findings, namely that the human eye moves considerably differently when reading real versus fake news, and that fake news detection can be roughly approximated using low-cost eye-tracking technology, are novel and of potential high impact to the technology in the area.
\end{abstract}

%\begin{CCSXML}
%<ccs2012>
%<concept>
%<concept_id>10002944.10011123.10010912</concept_id>
%<concept_desc>General and reference~Empirical studies</concept_desc>
%<concept_significance>500</concept_significance>
%</concept>
%<concept>
%<concept_id>10002951.10003317.10003331</concept_id>
%<concept_desc>Information systems~Users and interactive %retrieval</concept_desc>
%<concept_significance>500</concept_significance>
%</concept>
%<concept>
%<concept_id>10003120.10003121.10011748</concept_id>
%<concept_desc>Human-centered computing~Empirical studies in HCI</concept_desc>
%<concept_significance>100</concept_significance>
%</concept>
%</ccs2012>
%\end{CCSXML}

%\ccsdesc[500]{General and reference~Empirical studies}
%\ccsdesc[500]{Information systems~Users and interactive retrieval}
%\ccsdesc[100]{Human-centered computing~Empirical studies in HCI}

%\keywords{Factuality detection, eye tracking, fake news}

\maketitle
%\cl{General comments: (1) against my prior advice about paper writing, let us describe the eye-tracking experiment in the past tense. But its analysis and the rest of the paper in the present tense. (2) Let us not use both image and screen interchangeably. Pick one and use it consistently.}

%\cl{
%\begin{itemize}
%\item CH: GIVE DATA IN 2 FILES (AS PER BL'S EMAIL) TO JGS+BL+CL (14 JAN). WRITE SECTION 4 (18 JAN) 
%\item JGS: POLISH SECTION 2 (18 JAN), FEEDBACK SECTION 3 (16 JAN), WRITE SECTION 5 (22 JAN)
%\item BL: FEEDBACK SECTION 3 (16 JAN), POLISH SECTION 4 (22 JAN)
%\item CL: WRITE ABSTRACT (21 JAN), WRITE SECTION 1 (21 JAN), FEEDBACK SECTION 3 (16 JAN), POLISH SECTION 4 (22 JAN), WRITE SECTION 6 (22 JAN)
%\end{itemize}
%}
%\input{samplebody-conf}
\section{Introduction and Prior Work}
\label{s:intro}
%\cl{CL TODO BY 21 JAN}

%\begin{figure}
%    \centering
%    \includegraphics[width=0.5\linewidth,trim={0.2cm 1.2cm 2cm 1.2cm},clip]{images/r_4HEATMAP.jpg}}&
%    \includegraphics[width=0.5\linewidth,trim={0.2cm 1.2cm 2cm 1.2cm},clip]{images/r_5HEATMAP.jpg}}    
 %   \caption{}
 %   \label{fig:correlation}
%\end{figure}

%Motivated by the finding that, on many occasions, fake news tend to attract more overall clicks than real news \cite{Vosoughi1146}, w
%We study to what extent automated fact checking can be done based solely on eye-tracking data that is logged when users read news headlines.
%The dissemination of \textit{fake news} on the web, and its considerable political, financial and societal implications, have precipitated sharp criticism to web search engines and social network platforms for not containing or filtering out misinformation. One problem stems from the fact that, while search technology tends to approximate user interest and content relevance from user clicks, on many occasions fake news tend to attract more overall clicks than real news \cite{Vosoughi1146}. When this happens, misinformation spreads faster. This has stimulated research in automated fact checking, a.k.a. \textit{factuality checking}, where the state of the art uses deep learning architectures, for instance LSTMs, trained on labelled evidence and/or contextual data, which are typically also mined from the web \cite{C18-1283}. We investigate to what extent automated fact checking can be done based solely on eye-tracking data that is logged when users read news headlines.
Factuality detection in headlines is important because headlines are often solely responsible for the user's first impression (especially in mobile environments); but it is also challenging because, unlike full text, news headlines convey information succinctly and without reasoned argumentation or background. 

We measure the overt attention %when reading news headlines. According to the eye-mind assumption \cite{JustCarpenter80}, fixations, which occur when the eye temporarily rests on a stimulus (e.g. a word or phrase), correspond to the level of attention given to the corresponding information. During fixation time, the incoming information is processed, which is a sign of elaborate cognitive labour during reading comprehension. This leads to fixations being positively correlated to higher complexity and informativeness \cite{fraser-EtAl:2017:EMNLP2017} of the fixated stimuli. 
%We conduct a laboratory experiment where 
of 55 participants who are eye-tracked when reading 108 news headlines. %(crawled from the web, cf. Section \ref{ss:partmat}). We ask: [RQ1] Is human attention (measured through eye-tracking) distributed differently across headlines of true versus false news? [RQ2] How well can the factuality of a news headline be inferred using solely eye-tracked signals? 
We find statistically significantly longer eye gazing and fixation durations when reading headlines of true, rather than false news, regardless of participant gender. %(total gazing duration 3166.92ms vs 2770.37ms for true vs false headlines, fixation duration 807.57ms vs 677.00ms for true vs false headlines). 
%This finding is invariant with respect to gender.
We also train an ensemble learner, solely on eye-tracking data, to infer factuality in headlines. Our model yields a mean AUC of 0.688 and is better at detecting false headlines than true headlines. Further analysis shows that eye-tracking 25 users when reading 3-6 headlines is sufficient for our ensemble learner.
Eye tracking has long been used in IR to infer relevance \cite{ LobodaBB11, BuscherDE08,Buscher:2012:ADE:2070719.2070722, Hardoon:article,AjankiHKPS09,PuolamakiAK08} and to 
improve user understanding, for instance that adding information to search engine snippets significantly improves performance for informational tasks but degrades performance for navigational tasks \cite{CutrellG07}; %\chh{The previous description of the reference does not mention how they use eye-tracking} %found that adding information to search engine snippets significantly improved performance for informational tasks but degraded performance for navigational tasks. More recently, \citet{BhattacharyaG18} eye-tracked users searching the web and measured their topical knowledge before and after each task, finding 
that users with higher change in knowledge differ significantly in terms of the number and duration of fixations compared to users with lower knowledge-change \cite{BhattacharyaG18}; and that relevant documents tend to be continuously read, while irrelevant documents tend to be scanned \cite{Gwizdka14}. % studied users searching for information in news stories, finding that relevant documents tended to be continuously read, while irrelevant documents tended to be scanned. 
In most cases, cognitive effort inferred from eye-tracking data is highest for (at least) partially relevant documents and lowest for irrelevant documents. 

Our findings complement prior findings that news posts from credible sources receive more gaze attention \cite{SulflowSW19} and that false news tend to be read more quickly than accurate news \cite{Gwizdka14}. 
%The work of \citet{Gwizdka14} is compatible with our finding that false news tend to be read faster than factually accurate news. %We also verify the findings of \citet{WangYLCM14} when we observe no significant difference in number of fixations and fixation duration between the tasks of i) detecting the most recent news and ii) labelling news as false or true. \ch{I don't follow this point, we do not compare these two tasks directly, we do not test this. The only thing we know is that when true headlines are read they have more fixation and longer total fixation}
%\cl{This should also go here: }
%Our findings are also aligned with those by \citet{SulflowSW19}, who eye-track the reading of political news posts on Facebook by 103 participants, and find that news from more credible sources are read more often and for a longer time than news from less credible sources. %These findings are aligned with our findings that users spend more time reading true headlines than false headlines.
However, none of the above studies is done on headlines, and, to our knowledge, we present the first factuality inference model to be trained exclusively on eye-tracked data.
 
%\section{Eye movements as predictors of factuality}

%\jgs{This is a new (2019-01-16) section. It will describe background and the overall hypotheses (currently it contains material only from the experimental section). Depending on space issues, it should be merged into the introduction.}

%Hypothesis 1: fixation durations and saccade durations are overall longer when reading fake news statements than when reading true news statements. 

%Hypothesis 2: Eye fixation duration and saccade duration is a more accurate indicator of fake news detection, than the actual labels humans themselves give to the statements consciously when asked to do so.

\section{Experiment design}
\label{s:exp}
55 participants with normal or corrected-to-normal vision were recruited (24 females, 31 males; 19-33 years of age, median age 24),
and each participated in a single eye tracking session in a laboratory. 
At the start of each session, we logged the age and gender of each participant and then introduced the task and apparatus. The eye tracker was calibrated and the task commenced. On completion of the task, participants were debriefed and comments were solicited.  At no time were participants informed about how well they were doing. %At the end of the experiment, each participant was debriefed regarding the purpose of this research and comments were solicited. %\cl{Did some of them ask at the end?} \ch{A few but not many}
%
%\subsubsection{Task 1} %Initially the participants were only aware of the first task. 
Each participant was shown a screen (white background) with three headlines (each on a separate line, in black font, size=36), without any further information. The headlines were %positioned one under the other, 
centered on the screen, with 70mm of space between them and 20mm of space to the left border of the screen. Participants were asked to choose the most recent headline. This task was chosen on purpose to keep participants engaged in reading under circumstances where they were not directly checking for factuality. When participants had made their choice, the next screen (showing three new headlines) appeared. Participants did not know that two of the headlines were true and one was false, at any time. In total, 36 screens, each with three different headlines, were shown (108 unique headlines). To address order effects, we fully counterbalanced the position (top, middle, bottom) of the headlines, so that each position contained a factually false headline exactly 12 times. 
Participants could not move on to the next screen before answering, with no possibility of giving a ``don't know''-answer, %, and were asked to answer to the best of their ability. 
and could not revisit a previous screen. All participants saw the same 36 screens with the order of screens randomized across participants. %Eye movement was recorded throughout. 
%\subsubsection{Task 2}
%In the second task, participants were shown the exact same screens and headlines as in Task 1 (in random order), and were asked to label each headline as factually true or false. Participants were not informed on how many and which of the headlines were true or false. %Eye movements were recorded throughout this task too, but we do not use the eye-tracking data from this tas. Participants were presented with the same screens as in task 1, and were asked to classify each headline as either factually true or false.  %that each screen only contained a single factually wrong headline, and were asked to specify for each headline if it was factually true or false.
%\end{description}
No time limit was set for completing the task. 

To calibrate the experimental design, we did a pre-study on 11 participants with a subset of 24 screens. The pre-study did not lead to any changes in the design or protocol, except that the number of screens was increased to 36 because participants were faster than initially expected. In our analysis we combine the data from the pre-study with the remaining data to form the complete dataset.

Each participant performed the task individually, and was given the same oral instructions by the research assistant\footnote{\label{note1}\url{https://github.com/Varyn/Factuality_Checking_News_Headlines_EyeTracking}}. %all sessions were performed by the same research assistant. All participants received the  %(see the research protocol in Appendix \ref{sec:app_protocol}). 
Participants could at all times elect to stop the experiment (none did). The study was approved by the ethics board of our university, and all data was anonymized prior to storage and analysis.

%\paragraph{Participants and materials}
%\label{ss:partmat}
%\subsubsection{Participants}
%60 participants with normal or corrected-to-normal vision were recruited using social media and flyers at local universities. For five of them, the calibration of the eye tracker could not be completed\footnote{Unsuccessful calibration is to be expected in some instances, as e.g. different severities of strabismus provide a technical challenge for the eye tracker.}, so they were removed from the experiment, resulting in 55 participants (24 females, 31 males). Their age ranged from 19 to 33 years with a median age of 24. %No other demographics than age and gender were recorded. 
%Each participant was paid in kind with two cinema tickets. %for participating in the experiment. %, which on average took 45 minutes including briefing and debriefing.

%\cl{Did any of our participants wear glasses/lenses or had vision issues?} This did not consistently affect the calibration.
%\paragraph{Headlines and factuality}
The headlines shown to participants were crawled from the website of a reputable local newspaper\footnote{\url{https://www.thelocal.dk/}} and consisted of the full title of an article concerning \textit{local and national news}. From the pool of crawled headlines, we selected 108 headlines that:
(a) covered news that should be generally known to the public, (b) were formulated in approximately the same tone (i.e., no clickbait titles, no emphatics, no puns), and (c) were unlikely to provoke strong feelings. 
All headlines were selected manually by one of the authors of this paper (see Table \ref{tab:dataset} for their statistics). 
%used in this experiment were acquired using the headlines from a local newspaper, with the headlines chosen at random, with the requirement that the headline had to contain something which could be factually true or false.
%\jgs{This is almost certainly not how it went down. Christina should probably rewrite the previous headline to bring it in line with what she actually did.} %was considered factually true. \jgs{Did we corroborate with third party sources? How did we ensure factuality?} 

All crawled headlines were factually true. We created factually false headlines by semantically reversing parts of some headlines. For example, $\cdots$ \texttt{among \textbf{most} expensive cities to relocate to} became $\cdots$ \texttt{among \textbf{least} expensive cities to relocate to}. %(see examples in Table \ref{tab:false}). 
All the semantic transformations we used to falsify headlines are shown in Table \ref{tab:transf}. %For instance, the true headline ''\textit{SAS among world's most safe airlines}'' was falsified and became ''\textit{SAS among world's least safe airlines}''. %Other falsified examples and their origin can be seen in Table \ref{tab:false}. 
When falsifying headlines, we made sure that they still appeared semantically plausible and sounded natural. To make sure that there is no bias stemming from the linguistic formulation of true versus false headlines, we POS-tagged all headlines (using the Stanford parser) and found that the proportion of content words (which are known to be fixated on by the human eye much more than functions words \cite{Rayner98}) was approximately the same in both true and false headlines (see Table \ref{tab:dataset}). We make all 108 headlines freely available\textsuperscript{\ref{note1}}.

%\footnote{See footnote \ref{note1}}.

%\begin{table}[]
%    \centering
%    \scalebox{0.9}{
%    \begin{tabular}{lc}
%        \toprule
%        Headlines & 108 \\
%        True headlines & 72 \\
%        False headlines & 36 \\
%        Mean headline length & 55.32 \\
%        Minimum headline length & 29 \\
%        Maximum headline length & 88 \\
%        Avg. content words per true headline & 4.79 \\ 
%        Avg. content words per false headline & 4.53 \\
%        Avg. function words per true headline & 3.88 \\
%        Avg. function words per false headline & 4.08 \\
%        \bottomrule
%    \end{tabular}
%    }
%    \caption{Statistics of the 108 headlines (length measured as number of characters).}
%    \label{tab:dataset}
%\end{table}

\begin{table}[]
    \centering
    \caption{Dataset statistics.} %True (2/3) and false (1/3) headlines is 2/3 and 1/3 respectively, as per our experimental design.} %\cl{Please add headline length in words, not characters (otherwise the content/function words comparison could be misleading to the hasty reviewer)}
    \vspace{-9pt}
   \scalebox{0.8}{
    \begin{tabular}{lccc}
        \toprule
                        &True &False & Total \\ \midrule
        \# Headlines    &72 &36  &108 \\
        Mean \# words per headline & 8.56& 8.42& 8.51 \\
        Mean \# content words per headline & 4.79 &4.53& 4.70  \\ 
        Mean \# function words per headline & 3.88 &4.08& 3.95\\ 
        \bottomrule
    \end{tabular}
    }
    \vspace{3pt}
    \label{tab:dataset}
%\end{table}
%\label{false}
%\begin{table}[]
%    \centering
    \caption{All transformations that falsified news headlines.} %Several of these are symmetric (e.g. \textit{most} $\longrightarrow$ \textit{least} and \textit{least} $\longrightarrow$ \textit{most}).}
    \vspace{-9pt}
    \scalebox{0.8}{
    \begin{tabular}{ll}
        \toprule
         original text&transformed text\\
        \hline
more, most, best, top, highest, good&fewer, least, worst, bottom, lowest, bad\\
%&\\
%worst&best\\
%&bottom\\
%highest&lowest\\
denies, fear, pick up award, react to &admits, love, stripped of award, praise \\
%&\\
% & \\
two ... in top 50, remain, helping out&no ... in top 50, exit, refuses to help\\
%&\\
criticised, leads in, drops down&praised, last in, tops\\
%&\\
% &exit \\
%&admits\\
%&\\
cannot get enough of, calls for end&do no like, tolerates\\
%more&fewer\\
looks to $\ldots$ as inspiration&uses $\ldots$ as example to avoid\\
%&\\
%good&bad\\
%&&\\
\bottomrule
    \end{tabular}
    }
    \label{tab:transf}
    \vspace{-10pt}
\end{table}

%\noindent \textbf{Apparatus}.
%\subsection{Apparatus}
\paragraph{\textbf{Apparatus}}
\label{ss:apps_mats}
We used an Eyetribe ET1000 desk-mounted stream-based eye tracker bar, %(which was recently recommended for scientific research (see \citet{OomsK18}, Table 1). The eye-tracker was 
paired with a 24-inch screen (resolution of 1920x1200 and 170 DPI). The eye tracker sampled the position of eyes at the rate of 30 Hz and had a spatial resolution of 0.1 degree. 
We used iMotions\footnote{https://imotions.com/} to calibrate the eye tracker and collect the data. Participants were placed 60cm away from the screen, and the room had soft standard artificial light. No head stabilisation was used (head movements were unconstrained so the intrusion of the eye moving measurement was minimal). %Seating adjustments were made to account for differences in participant height to ensure proper alignment of the eye-tracker optics. 
We calibrated the eye tracker using a standard 9-point calibration prior to each recording. %An integrated log of eye-movement data, facial data, and factuality labels allowed us to map eye movements to various features on the screen during the performed tasks. %\cl{And facial expressions, right?}

Participants indicated which of the three headlines per screen was the most recent by typing 1, 2, or 3 on the keyboard (for the position of the top, middle, and respectively bottom headline). Typing was chosen over using the cursor because the cursor could interfere considerably with eye tracking. %In Task 2, participants used the cursor to click on check buttons at the bottom of the screen to label the factuality of each headline. Even though we collected eye-tracking data from both tasks, in this paper, we use only the eye-tracking data collected during Task 1. We do not use the data from Task 2 because, in addition to cursor interference, participants were already familiar with the headlines \textit{and} were consciously thinking about their factuality (whereas we wanted to eye-track them when confronted with unseen headlines and unaware of any potential factuality scenario). %and the cursor interfered with the eye-tracking%The mouse was not used for Task 1, so that it would not interfere with the eye tracking, as the cursor  
%\cl{Surely that must have happened in task 2, and in prior work, right? Can we cite some? Why did we decide not to care about this in task 2? It seems inconsistent.} \ch{This was a limitation in the software, which customer service said would be fixed in a future version.}
%\jgs{Did they click directly on the headline with the mouse pointer, or on something else? Also, how do we explain why we only gave them a mouse for Task 2?} 

%\noindent\textbf{Eye tracking measures}.
% \subsection{Eye tracking measures}
\paragraph{\textbf{Eye-tracking measures}}
%listed below, which are standard eye-tracking measures \cite{Jacob2003-JACETI}, consistently measurable with high precision in most off-the-shelf commercial eye-tracking apparatus. %(as opposed to more intricate movements such as pupil dilation or saccades \cl{is this a risky claim?}). 
A fixation is a stable eye-in-head position within a dispersion threshold (typically 2 degrees), above a duration threshold (typically 100-200 milliseconds\footnote{We set fixations at 100 milliseconds.}), and velocity below a threshold (typically 15-100 degrees per second).
Gaze duration is the cumulative duration of a sequence of consecutive fixations within an area of interest (AOI). We defined a separate AOI around each headline and we analysed these 5 measures: %We used a font size of 36 for each headline, and defined the AOI to be approximately one line above and below the headline text.
the total time spent fixating inside an AOI (\textbf{total fixation duration}); the total number of fixations inside a AOI (\textbf{total fixation count}; the total time spent gazing inside an AOI (\textbf{total gaze duration})\footnote{Gaze duration consists of the duration of fixations and other captured gaze activity (such as time between fixations) inside an AOI.}; the \textbf{average fixation duration} inside an AOI (total fixation duration divided by total fixation count); the duration of the first fixation inside an AOI (\textbf{first fixation duration}).

\section{Findings}
\label{stat_design_rq1}

%We next present how we investigate the link between eye movement and headline factuality (RQ1). %\cl{Add facial expressions.}

%\paragraph{RQ1 (Eye Movement and Headline Factuality)}
We now study the statistical effect the headline factuality has on the eye-tracking measures.
Let $\gamma$ denote any of the above 5 eye-tracking measures. %: \textit{total fixation duration, total fixation count, total gaze duration, average fixation duration,} and \textit{first fixation duration}. Each of these may be possibly influenced, not only by factuality, but also by the length of the headline, the gender of the participant, and the position of the headline in the screen. 
To establish whether factuality affects each of these $\gamma$s in a statistically significant way, we consider both fixed effects (gender, headline length, position of headline on screen), and random effects. These fixed and random effects are potentially non-negligible, meaning that conventional methods for inferential data analysis, such as ANOVA and general linear regression are not applicable \cite{LobodaBB11}. We therefore fit a mixed model \cite{mixedcomplex} that uses the above $\gamma$s as a response and the fixed effects as explanatory variables. Because each participant is drawn from some larger population, the participant is included as a random intercept. 
%We have data for only 55 participants, so we only consider a first order model, and do not include any interaction between the factors. %\cl{Such as what?} \ch{Did this clarify?} \ch{this line is not important, we have the following model, which is very standard. Stating that we only consider first order because of data easily open us up to people saying, "I think we could do a higher order model with that amount of data".}
The mixed model for each of the above $\gamma$s is:
\begin{equation}
\label{eq:mixed_model}
    \gamma = c_{\textrm{true}}i_{\textrm{true}} + c_{\textrm{middle}}i_{\textrm{middle}} + c_{\textrm{bottom}} i_{\textrm{bottom}} \nonumber  
    + c_{\textrm{male}}i_{\textrm{male}} + c_{\textrm{length}} l 
    + p + b
\end{equation}
%\jgs{@Christian, Christina: Are you ok with the linebreaks in the equation above?} \ch{Totally ok, was to long before}\cl{OK looks good}
\noindent where $c_{\textrm{factor}}$ is the coefficient for the factor and
 $i_{\textrm{factor}}$ is the indicator function for the factor, e.g. $i_{\textrm{male}} = 1$ if the participant is male and $i_{\textrm{male}} = 0$ otherwise. For the categorical variables of position (middle, bottom), gender (male), and factuality (true), there are $k-1$ fewer factors than number of categories ($k$). $l$ is the normalised length of the headline with zero mean and unit variance, $p$ is the random effect for the participant, and $b$ is the intercept. The model is fitted using the $\gamma$s collected; these $\gamma$s are normalised so that the scale of the coefficient is comparable across measures, which otherwise have different scales.

The coefficient $c_{\textrm{true}}$ shows the relation between the measure \(\gamma\) and the factuality of the headline. We formulate the null hypothesis $H^{\gamma}_0$ for $\gamma$ as the assumption that factuality does not affect $\gamma$, that is 
$H^{\gamma}_0 : c_{\textrm{true}} = 0$. To test this hypothesis, we compute $p$-values and confidence intervals for each coefficient by performing Wald tests. We have 5 different eye-tracking measures, so we perform 5 hypothesis tests with Bonferroni correction, requiring that $p < \frac{0.05}{5}=0.01$ to reject each 
$H^\gamma_0$.
All statistical analysis is done using StatsModels\footnote{\url{https://www.statsmodels.org/stable/index.html}, version 0.9}, and the models are fitted using Maximum Likelihood.

%\subsection{Results and Analysis}
%\label{s:ana}
%\cl{CH TO WRITE BY 18 JAN. CL+BL TO POLISH BY 22 JAN}
%We present the results and analysis for each of the research questions of the eye tracking study.
%Before these analysis are done, we will first present the measures related to fixation and saccade duration which are used in this project. Following this some basic statistics of the collected data will be presented, and lastly the analysis for verifying hypothesis 1 and 2 will be presented.

%\subsection{Statistics of the eye tracked data}
%\label{ss:basic_stat}
%In this section we will shortly present the basic statistics of the collected data.

%\paragraphs{Results}
%\label{ss:rq1}
%We investigate the coefficients for factuality for each of the five eye-track measures (see Section \ref{ss:apps_mats}). 
Table \ref{tab:answer_stat2} shows the resulting coefficients. We see that for \textit{total gaze duration, total fixation duration}, and \textit{total fixation count} %the coefficient
%$c_{\textrm{true}} > .1$ %and 
$p < .001$, thus we have sufficient evidence to reject the null hypothesis. These three eye-tracking measures change significantly when reading true versus false headlines. However, for \textit{average fixation duration} and \textit{first fixation duration}, we cannot reject the null hypothesis, and thus we cannot conclude that the time spent on each individual fixation changes between factually true and false headlines. %Inspecting the coefficients for the three statistically significant measures affected by factuality, w
We also observe that a factually true headline causes the \textit{total gaze duration, total fixation duration}, and \textit{total fixation count} to increase, as seen by the positive value of $c_{\textrm{true}}$; this means that false headlines in general have shorter fixation and gazing duration than true headlines. The fact that factuality is not significant for \textit{average fixation duration} means that the increased \textit{total fixation duration} for true headlines is caused by an increase in \textit{total fixation count} for factually true headlines.

%The factuality of the headline has the largest coefficient of the model which predicts "total gaze duration", but it is not significantly larger than the coefficient for the two other measures, as the confidence interval of the coefficients overlaps. \ch{I do not like this observation, is a bit "bad statistics", trying to find another point}

%The coefficients for the fixed effects of the fitted models for all five measures in Table \ref{tab:answer_stat2} show the same tendency for all non-factuality factors independently of the measure: gender is insignificant, longer headlines in general have a higher score, and headline position is significant ( scores get lower moving from the top to the bottom position).
%We briefly discuss the coefficients for the fixed effects of the fitted models for \textit{total gaze duration} (a similar discussion could be had for the other measures as they all show the same tendencies in Table \ref{tab:answer_stat2}). 
We now briefly discuss the other coefficients than $c_{\text{true}}$. 
Using $p<0.01$, we see that the position of the headline is not significant for the \textit{total gaze duration}, while it is significant if the headline is placed on the bottom for all measures of fixation. The negative value of $c_{\text{bottom}}$ shows that all measures of fixation decrease when the headline is placed on the bottom. The length of the headline is significant for all eye-tracking measures (p < 0.001), with longer headlines having higher measures. Lastly, we observe no significant difference in any measures between the genders.

\begin{table}[]
    \centering
        \caption{The fixed effects for the five eye-tracking measures. $p$-values below 0.01 are marked in bold. %The intercept is not listed as it is uninteresting. 
    (See Section \ref{stat_design_rq1} for notation).
    }
    \vspace{-7pt}
    \scalebox{0.76}{
    \begin{tabular}{lcccccc}
        \toprule
       &             Coef. &  Std.Err. &  z &   P>|z| & [0.025 & 0.975] \\ \hline
\textbf{Total gaze duration} & \\ \hline
\(c_{\textrm{true}}\)&  0.154&    0.023&  6.697& \textbf{<0.001} &  0.109&  0.199 \\
\(c_{\textrm{middle}}\) &    -0.026&    0.027& -0.959& 0.338& -0.078&  0.027 \\
\(c_{\textrm{bottom}}\) &    -0.054&    0.027& -2.020& 0.043& -0.106& -0.002 \\
\(c_{\textrm{male}}\) &      -0.149&    0.154& -0.969& 0.333& -0.451&  0.153 \\
\(c_{\textrm{length}}\)&               0.174&    0.011& 15.844& \textbf{<0.001} &  0.153&  0.196 \\
\hline
\textbf{Total fixation duration} & \\ \hline
\(c_{\textrm{true}}\)&  0.109&    0.021&   5.301& \textbf{<0.001} &  0.069&  0.149 \\
\(c_{\textrm{middle}}\) &    -0.083&    0.024&  -3.474& \textbf{<0.001} & -0.129& -0.036 \\
\(c_{\textrm{bottom}}\) &    -0.239&    0.024& -10.059& \textbf{<0.001} & -0.285& -0.192 \\
\(c_{\textrm{male}}\) &      -0.202&    0.182&  -1.109& 0.267& -0.558&  0.155 \\
\(c_{\textrm{length}}\)&               0.100&    0.010&  10.154& \textbf{<0.001} &  0.081&  0.119 \\
\hline
\textbf{Total fixation count} & \\ \hline
\(c_{\textrm{true}}\) &  0.115&    0.020&  5.609& \textbf{<0.001}&  0.075&  0.155 \\
\(c_{\textrm{middle}}\) &    -0.037&    0.024& -1.536& 0.124& -0.083&  0.010 \\
\(c_{\textrm{bottom}}\) &    -0.199&    0.024& -8.420& \textbf{<0.001}& -0.246& -0.153 \\
\(c_{\textrm{male}}\) &      -0.164&    0.184& -0.894& 0.371& -0.524&  0.196 \\
\(c_{\textrm{length}}\)&               0.118&    0.010& 12.011& \textbf{<0.001}&  0.099&  0.137 \\
\hline
\textbf{Average fixation duration} & \\ \hline
\(c_{\textrm{true}}\)&      0.025&      0.022&    1.106&   0.269&   -0.019&    0.068 \\
\(c_{\textrm{middle}}\) &        -0.003&      0.026&   -0.125&   0.900&   -0.054&    0.047 \\
\(c_{\textrm{bottom}}\) &        -0.130&      0.026&   -5.061&   \textbf{<0.001}&   -0.181&   -0.080 \\
\(c_{\textrm{male}}\) &          -0.006&      0.171&   -0.038&   0.970&   -0.342&    0.329 \\
\(c_{\textrm{length}}\)&                   0.059&      0.011&    5.509&   \textbf{<0.001}&    0.038&    0.079 \\
\hline
\textbf{First fixation duration} & \\ \hline
\(c_{\textrm{true}}\) &      0.034&     0.024&   1.411&  0.158&  -0.013&   0.081 \\
\(c_{\textrm{middle}}\) &          0.014&     0.028&   0.484&  0.628&  -0.041&   0.068 \\
\(c_{\textrm{bottom}}\) &         -0.120&     0.028&  -4.321&  \textbf{<0.001}&  -0.175&  -0.066 \\
\(c_{\textrm{male}}\) &           -0.016&     0.148&  -0.106&  0.915&  -0.305&   0.274 \\ 
\(c_{\textrm{length}}\)&                    0.056&     0.011&   4.906&  \textbf{<0.001}&   0.034&   0.079 \\
        \bottomrule
    \end{tabular}
    }
    \label{tab:answer_stat2}
    \vspace{-10pt}
\end{table}

\paragraph{\textbf{Learning to infer factuality from eye tracking}}
Having established that \textit{total gaze duration, total fixation duration}, and \textit{total fixation count} are all significantly different depending on the headline factuality, we next investigate if these measures provide sufficient signal for training a headline factuality classifier. As these measures are highly dependent on the length and position of the headlines, they are also included in the model. We observe that \textit{total fixation duration} is highly correlated with \textit{total fixation count}, thus to keep the model as simple as possible, we only use \textit{total gaze duration} and \textit{total fixation duration}.

In table \ref{tab:answer_stat2}, we see the coefficient of factuality ($c_{\textrm{true}}$), for many measures, to be less influential than the position and length of the headline. Thus, we expect using eye-tracking measures of only a single participant to be noisy. Due to this, we use an ensembling approach, where the predicted factuality of a headline is computed as an average over a set of participants ($P_{\textrm{ens}}$): 
$
    v_h = \frac{1}{|P_{\textrm{ens}}|} \sum_{p \in P_{\textrm{ens}}} v_{p,h}
$,
where $v_h$ is the factuality prediction for headline $h$, and $v_{p,h}$ is the factuality prediction for headline $h$ for participant $p$. 
Due to the relative small size of our dataset, we propose to use the average of two simple second-order logistic models for estimating $v_{p,h}$:
\begin{align}
    \label{eq:v1}v^1_{p,h} &= \frac{1}{1+e^{-(c_1i_{\textrm{top}}(h)\gamma^{\textrm{GD}}_{p,h}+c_2i_{\textrm{middle}}(h)\gamma^{\textrm{GD}}_{p,h}+c_3i_{\textrm{bottom}}(h)\gamma^{\textrm{GD}}_{p,h})}} \\
    \label{eq:v2}v^2_{p,h} &= \frac{1}{1+e^{-(c_4 l_ h\gamma^{\textrm{FD}}_{p,h})}}, \;\;\;\; v_{p,h} = \frac{v^1_{p,h}+v^2_{p,h}}{2}
\end{align}
where $c_k, k \in [1,2,3,4]$ are the learned coefficients of the logistic models, $\gamma^{\textrm{GD}}_{p,h}$ is the total gaze duration for participant $p$ on headline $h$, $\gamma^{\textrm{FD}}_{p,h}$ is the total fixation duration for $p$ on $h$, and $l$ is the length of the headline. Both logistic models have one eye-tracking measure interacting with either the length or position of the headline, where the interaction is chosen based on the pair with the lowest correlation. We choose to use two simple logistic models, instead of a single combined model, to increase the variance of the predicted factuality, as high variance is beneficial for ensembling. We standardize (zero mean and unit variance) the eye-tracking measures from each participant across all headlines. Lastly, the two logistic models are trained using Maximum Likelihood on a set of training participants.

\paragraph{\textbf{Evaluation}}
We evaluate the model by inferring factuality on unseen headlines using Monte Carlo cross-validation over 100,000 iterations. In each iteration, the participants are split for training and ensembling (27 and 28 participants, respectively), and three headlines are chosen for evaluation (2 true and 1 false), while the remaining headlines are used for training. We report the mean AUC and mean accuracy, across all iterations.

%We split the dataset so that 27 participants are used to train the model and 28 participants are used for testing (i.e., part of the ensemble set \(P_{\textrm{ens}}\)). We train the model on 105 headlines, and test the trained model on 3 unseen headlines. We use stratified sampling such that the 3 unseen headlines always consist of 2 true and 1 false headline. Due to the sampling of both the train/test participants, and the test headlines, we use Monte Carlo cross-validation with 100000 iterations. We report the mean AUC and mean accuracy, across all iterations.

As reported in Table \ref{tab:results}, we find that our ensemble model predicts the factuality of unseen headlines with a mean AUC of 0.688 and an accuracy of 0.634 (which is higher on false headlines (0.662) than one true ones (0.619)). There is no prior work on automatically detecting factuality in news headlines only, but related work on inferring factuality in text (but not headlines, which is harder) using textual features alone (not eye-tracking features), shows that accuracy ranges from 0.39 \cite{Wang17} to 0.76 \cite{Perez-RosasKLM18}, and even up to 0.86 \cite{poliak-EtAl:2018:S18-2} when using BiLSTMs and a multilayer perceptron classifier with refined linguistic features such as entailment and contradiction. Comparably, we have a simple learning model, which uses weaker input features (eye-tracking measures are less discriminative than textual ones), and which solves a more difficult problem (factuality checking in headlines instead of longer texts). 

\begin{table}[]
    \centering
    \caption{Factuality performance scores from our eye-tracking ensemble model.}
    \vspace{-6pt}
    \resizebox{\linewidth}{!}{
    \begin{tabular}{cccc}
    \toprule
        Mean AUC & Mean Acc. & Mean Acc. (True) & Mean Acc. (False) \\
        \midrule
        0.688 & 0.634 & 0.619 & 0.662 \\
        \bottomrule
    \end{tabular}}
    \label{tab:results}
    \vspace{-10pt}
\end{table}

\paragraph{\textbf{Analysis}}
In the above, we standardize the eye-tracking measures for each participant on all headlines. We now ask: how important is this standardization, and would standardization on fewer headlines suffice? We answer this by sampling fewer headlines to base the standardization on, while still preserving the ratio of 2 true headlines for each false one. We refer to three headlines following this ratio as a ``screen''.

\begin{figure}[]
    \centering
    \includegraphics[width=0.49\textwidth]{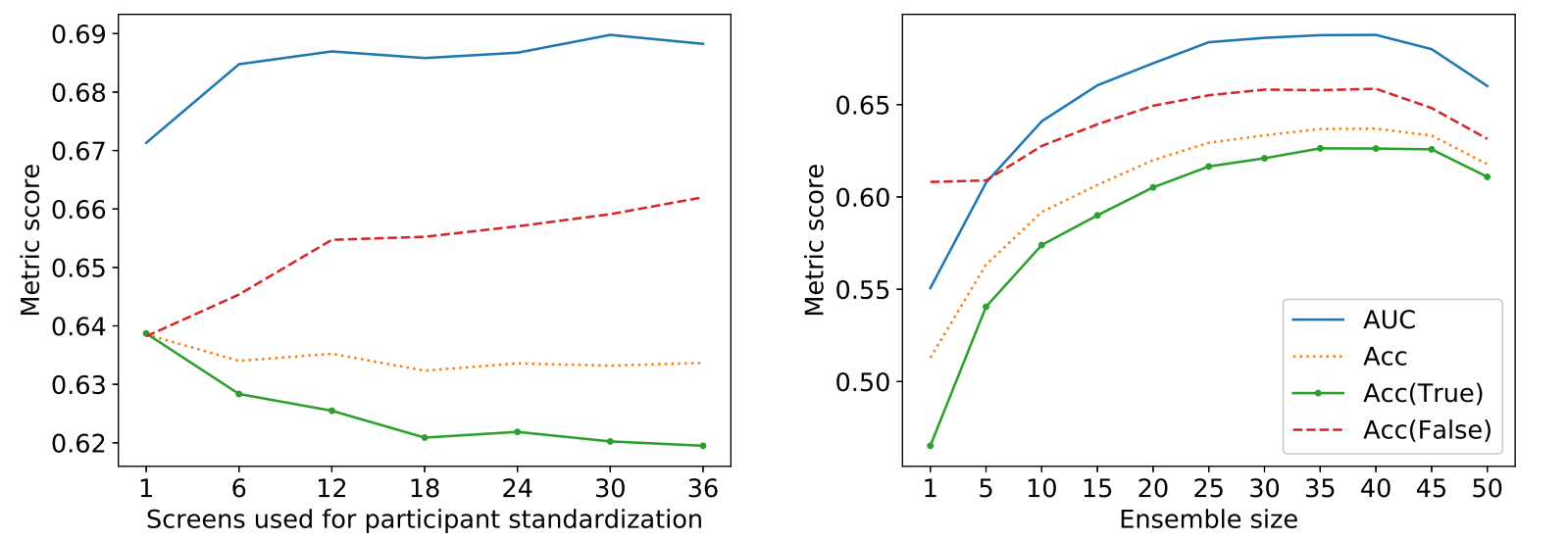}
    % \begin{subfigure}[b]{0.49\linewidth}
    %     \includegraphics[width=\textwidth]{images/subject_norm_new4.eps}
    %     %\caption{}
    %     \label{fig:rq2-participant-normalisation}
    % \end{subfigure}
    % \begin{subfigure}[b]{0.49\linewidth}
    %         \includegraphics[width=\textwidth]{images/grp_size_new4.eps}
    %     %\caption{}
    %     \label{fig:rq2-test-size}
    % \end{subfigure}
    %\begin{subfigure}[b]{0.33\textwidth}
    %        \includegraphics[width=\textwidth]{images/human_perf2.pdf}
%
 %       \caption{}
  %      \label{fig:human-perf}
   % \end{subfigure}
   \vspace{-20pt}
    \caption{Performance analysis when varying (left) the number of screens used for participant standardization in our model and (right) the number of participants used for ensembling.}\label{fig:rq2-analysis} 
    \vspace{-10pt}
\end{figure}

%\subsubsection{Effect of normalisation data} \label{sss:norm}

%\begin{figure}[h]
%    \centering
%    \includegraphics[width=0.6\linewidth]{images/subject_norm_new.pdf}
%    \caption{Performance analysis when varying the number of screen used for participant normalisation.}\label{fig:rq2-participant-normalisation}
%\end{figure}

Figure \ref{fig:rq2-analysis} (left) shows the mean accuracy and AUC when varying the number of screens used for standardization. When only standardizing on the screen we predict on (screen=1), mean AUC is at minimum; it drastically increases at 6 screens, and then stabilizes for the remaining number of screens. When increasing the number of screens, the accuracy for the true headlines decreases slightly, while the accuracy increases for the false headlines, but after 6-18 screens the difference of including more screens is minimal. 
This suggests that the performance of our ensemble model is not largely dependent on a large set of headlines to use for standardization. Deployed on a live setup, few headlines for standardization could suffice to fetch the accuracy and AUC levels reported in this study.

%The mean accuracy is marginally better when using a single screen compared to a larger amount, but generally the variation in accuracy across the range of screens is low. This indicates than mean accuracy is relatively stable (and hence unaffected) by the screen normalisation.
%This suggests that the performance of our inference model is not largely dependent on a large set of headlines to normalise on. Deployed on a live setup, few training headlines could suffice to fetch the accuracy and AUC levels reported in this study.

%\subsubsection{Effect of train-test split of eye-tracked participants} %\label{sss:split}
The results reported above correspond to splitting participants approximately 50/50 for training and ensembling, and this split can of course be varied; Figure \ref{fig:rq2-analysis} (right) plots mean accuracy and AUC (y axis) across a varying number of participants used for ensembling out of the 55 participants in total. We see that the choice of a 50/50 split is close to optimal. The fact that performance drops rapidly when 15 or fewer participants are used for ensembling indicates that aggregating over a large set of participants is at least as important as training a model on more data, in this setup. This happens because our dataset is small (we have few participants), so the optimal performance is a trade-off between training a better model (requiring more participants for training) and aggregating over more participants (requiring more participants in the ensemble). 

\section{Conclusions}
\label{s:conc}
%\cl{CL TODO BY 22 JAN}
We studied whether the human eye moves differently when reading factually true versus factually false news headlines, and if we can infer factuality in news headlines using only eye-tracking signals.
In an experiment with 55 users reading 108 news headlines, we found that false headlines receive statistically significantly less visual attention than true ones. We used this to build an ensemble learner that predicts news headline factuality using only eye-tracking measurements, which obtained a mean AUC score of 0.688 and a mean accuracy of 0.634. 

Future work includes investigation of eye tracking as a boosting mechanism to potentially improve factuality detection based on text processing, and refining the relationship between eye movements in more typical IR tasks such as search. A different direction of promising future work is to repeat our study 
  ``in the wild'' outside usual laboratory settings, including eye-tracking methods with lower fidelity, such as for instance typical cameras mounted on laptops and smartphone cameras.

%\clearpage

%\section{Headline falsification transformations}
%\label{ss:transformations}

%Table \ref{tab:transf} shows all semantic transformations used to falsify news headlines.
%\newpage
%\clearpage
\bibliographystyle{ACM-Reference-Format}
\bibliography{main.bib}

\end{document}